\documentclass[aps,pre,preprint,groupedaddress,showpacs]{revtex4-1}
\usepackage{amssymb}

\usepackage[dvips]{graphicx}
\usepackage{textcomp}
\usepackage{amssymb}

\newcommand{\mc}{\multicolumn}

\begin{document}

\title{ \Large\bf A Monte Carlo study of surface critical phenomena: 
       The special point}

\author{Martin Hasenbusch}
\email[]{Martin.Hasenbusch@physik.hu-berlin.de}
\affiliation{
Institut f\"ur Physik, Humboldt-Universit\"at zu Berlin,
Newtonstr. 15, 12489 Berlin, Germany}

\date{\today}

\begin{abstract}
We study the special point in the phase diagram of a semi-infinite system, where 
the bulk transition is in the three-dimensional Ising universality class.
To this end we perform a finite size scaling study of the improved 
Blume-Capel model on the simple cubic lattice with two different 
types of surface interactions. In order to check for the effect of 
leading bulk corrections we have also simulated the spin-1/2 Ising model on  
the simple cubic lattice. We have accurately estimated the
surface enhancement coupling at the special point 
of these models. We find $y_{t_1}=0.718(2)$ and $y_{h_1}=1.6465(6)$ 
for the surface renormalization
group exponents of the special transition. These results are compared
with previous ones obtained by using field theoretic methods and
Monte Carlo simulations of the spin-1/2 Ising model. Furthermore
we study the behavior of the surface transition near the special point
and finally we discuss films with special boundary conditions at 
one surface and fixed ones at the other.
\end{abstract}
\pacs{05.50.+q, 05.70.Jk, 05.10.Ln, 68.15.+e}
\keywords{}
\maketitle

\section{Introduction}
In this paper we shall study the special point in the 
phase diagram of a semi-infinite system.  For  reviews on surface critical
phenomena see refs. 
\cite{BinderS,Diehl86,Diehl97}. Let us briefly recall the basic
features of this phase diagram at the example of the spin-1/2 Ising
model on the simple cubic lattice and a semi-infinite geometry. 
It's reduced Hamiltonian is given by
\begin{equation}
\label{HIsing}
 H = - \beta \sum_{<xy>} s_x s_y -h \sum_{x}  s_x 
 - \beta_1 \sum_{<xy> \in S} s_x s_y -h_1 \sum_{x \in S}  s_x \;\;,
\end{equation}
where $x$ and $y$ denote sites of the lattice, $<xy>$ is a pair of nearest 
neighbors and $S$ is the surface of the system. The spin $s_x$ at the site $x$
can take either the value $-1$ or $1$. The Boltzmann factor is given by
$\exp(-H)$, since the reduced Hamiltonian incorporates the 
temperature. We define $\beta=J/k_B T$, $\beta_1=J_1/k_B T$, where $J$ is
the coupling constant in the bulk, $J_1$ the excess coupling constant at the 
surface and $T$ is the temperature. Below we shall refer to $\beta$ and $\beta_1$
as the coupling in the bulk and the excess coupling at the surface, respectively.  
Below we shall consider vanishing external fields $h=h_1=0$ in the bulk
as well as at the surface.

\begin{figure}[tp]
\includegraphics[width=13.3cm]{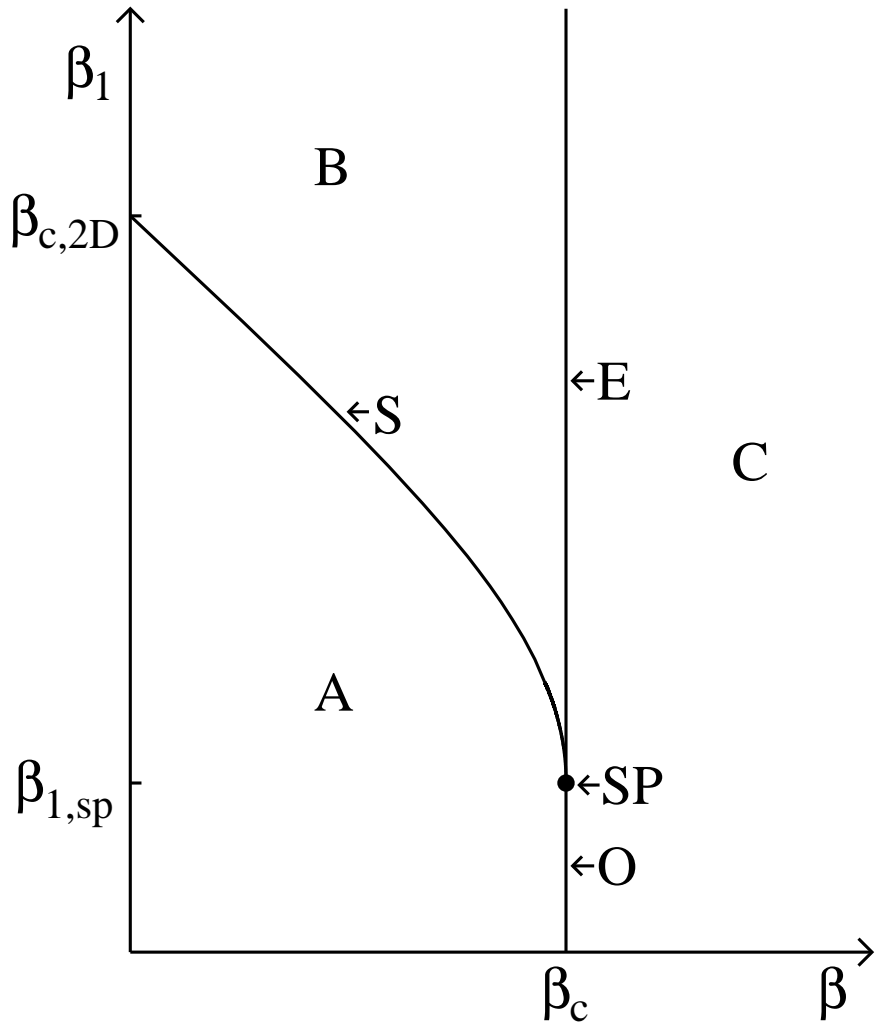}
\caption{\label{diagram}
Sketch of the phase diagram of the semi-infinite system. 
On the $x$-axis we plot the 
coupling $\beta$ of the bulk and on the $y$-axis the excess coupling $\beta_1$
of the surface. The phase, where both the boundary spins and those of the
bulk are disordered is labelled by A, while the one with disordered bulk and
and ordered surface is labelled by B. The phase, where the bulk and the 
surface are ordered 
is labelled by C.  The line of surface transitions is labelled by S, the 
line of ordinary transitions by O and the line of extraordinary transitions
by E. These three lines meet in the  so called special or surface-bulk point
that we have labelled by SP. A discussion is given in the text.}
\end{figure}

In figure \ref{diagram} we have sketched the phase diagram of this system.
For $\beta > \beta_c$, where $\beta_c$ is the critical coupling of 
the bulk system, the spins in the bulk are ordered. As a consequence,
also the spins at the surface are ordered. At 
$\beta=0$ the spins at the surface decouple completely from those of the bulk.
Hence a two dimensional Ising  model remains that undergoes 
a phase transition at $\beta_1=\beta_{c,2D}$. Starting from the point
$(0, \beta_{c,2D})$ there is the line of surface transitions, where the spins 
at the surface order, while those of the bulk remain disordered.
This line hits the line $(\beta_c,\beta_1)$ in the so called special
or surface-bulk point,
which is a tri-critical point. The transitions from disordered surface and
disordered bulk to ordered  bulk and ordered surface are called ordinary transitions,
while those from disordered bulk and ordered  surface to ordered bulk and 
ordered surface are called extraordinary transitions.

Surface critical phenomena had been studied first by using the mean-field 
approximation \cite{BinderS}. The application of  field theoretic methods to 
surface critical phenomena is complicated by the fact that translational 
invariance is broken by the surface. As a consequence, surface critical
exponents are only computed up to $O(\epsilon^2)$ in the $\epsilon$-expansion
\cite{Diehl86,Diehl97}. Furthermore surface critical phenomena have been 
studied by using high temperature series expansions, real space renormalization
group methods and Monte Carlo simulations of lattice models.  

The special point is characterized by the two relevant bulk renormalization group 
(RG)-exponents
$y_t$ and $y_h$ and the two relevant surface RG-exponents $y_{t_1}$ and 
$y_{h_1}$.  Similar to the pure bulk case, these  RG-exponents can be related 
to a number of surface critical exponents that characterize the behavior of 
thermodynamic quantities related to the surface in the neighborhood of the special 
point \cite{BinderS,Diehl86,Diehl97}.

In the present work we locate the special point of three different lattice 
models and determine the 
RG-exponents $y_{t_1}$ and $y_{h_1}$ by using finite size scaling (FSS)
\cite{Barber} methods.  In the presence of a surface, typically
corrections $\propto L^{-1}$ appear \cite{BinderS,Diehl86,Diehl97},
where $L$ is the linear extent of the
finite system. Analysing numerical data, it is difficult to disentangle
these corrections from leading bulk corrections which are
$\propto L^{-\omega}$, where $\omega=0.832(6)$ \cite{mycritical}.
Therefore, in addition to the Ising model  we simulate the
Blume-Capel model that is  a generalization of the Ising model.
In addition to $-1$ and $1$ the spin can take the value $0$. The parameter
$D$ of the Blume-Capel model controls the density of spins with $s_x=0$.
It has been shown that for a particular value $D^*$ of the parameter $D$ 
the amplitude of leading corrections to scaling vanishes.  See \cite{mycritical}
and refs. therein. The precise definition of the model is given below
in section \ref{definemodel}.  

The outline of the paper is the following: First we define the models 
that we simulate. We summarize results for the critical coupling 
and the critical exponents of the bulk system. We define the 
quantities  that we have measured and discuss their finite size 
scaling behavior. The scaling behavior is affected by corrections,
which have to be taken into account when analysing Monte Carlo data. 
Then we report
our numerical results: First we have simulated three different models
on $L^3$ lattices, where $L$ is the linear extend of the lattice. 
Based on these simulations we determine $\beta_{1,sp}$ for these models
and obtain estimates of $y_{t_1}$ and $y_{h_1}$. Next we study 
the surface transition in the neighborhood of the special point.
Then we discuss the magnetisation profile of films with special 
boundary conditions at one surface and fixed boundary conditions 
at the other. Finally we summarize and compare our results
for the RG-exponents with those obtained by field theoretic methods
and previous simulations of the spin-1/2 Ising model.

\section{Models and observables}
\label{definemodel}
The Blume-Capel model on the simple cubic lattice
is characterized by the reduced Hamiltonian
\begin{equation}
 H = -\beta \sum_{<xy>}  s_x s_y + D \sum_x s_x^2  - h \sum_x s_x \;\;,
\end{equation}
where $x=(x_0,x_1,x_2)$ denotes a site of the lattice. The components 
$x_0$, $x_1$ and $x_2$ take integer values in the range $1 \le x_i \le L_i$. 
The spin $s_x$ might take the  
values $-1$, $0$ or $1$. In the following we shall consider a vanishing 
external field $h=0$ throughout. The parameter $D$ controls the density of 
vacancies $s_x=0$. In the limit $D \rightarrow - \infty$, these vacancies
are completely suppressed, and hence the spin-1/2 
Ising model is recovered. For $-\infty \le  D < D_{tri}$ the model undergoes 
a second order phase transition in the three-dimensional Ising
universality class. For $D> D_{tri}$ the transition is of first order.
The most recent estimate for the tri-critical point is
$D_{tri} = 2.0313(4)$ \cite{DeBl04}. 
Numerically, using Monte Carlo simulations it has been shown that there
is a point $(D^*,\beta_c(D^*))$ on the line of second order
phase transitions, where the amplitude of leading corrections to scaling 
vanishes.  Our most recent estimate is $D^*=0.656(20)$ \cite{mycritical}.
In \cite{mycritical} we have simulated the model at $D=0.655$ close to 
$\beta_c$ on lattices of a linear size up to $L=360$. From a standard finite
size scaling analysis of renormalization group invariant quantities
such as the Binder cumulant we found
\begin{equation}
\beta_c(0.655)=0.387721735(25) 
\end{equation}
for the  critical coupling of the bulk at $D=0.655$. Recent
estimates for the critical coupling of the bulk of the spin-1/2 Ising
model are $\beta_c= 0.22165455(3)$, table X of \cite{DeBl03X}  
and $0.22165463(8)$  \cite{mycritical}. In the following we assume
$\beta_c= 0.2216546$.
The amplitude of leading corrections to scaling at $D=0.655$ is at least
by a factor of $30$ smaller than for the spin-1/2 Ising model. 

Our recent estimates for bulk critical exponents in the three-dimensional
Ising universality class are \cite{mycritical}
\begin{eqnarray}
 \nu &=& 0.63002(10) \;\;,\\
\eta &=& 0.03627(10) \;\;,\\
\omega &=& 0.832(6) \;\;.
\end{eqnarray}

\subsection{Film geometry and boundary conditions}
\label{definefilm}
In Monte Carlo simulations we are restricted to finite lattices. 
Therefore a surface requires a counterpart. This means that we actually 
study systems with a film geometry.
In the ideal case, film geometry means that the system has a finite
thickness $L_0$, while in the other two directions the thermodynamic
limit $L_1, L_2 \rightarrow \infty$ is taken. In order to approximate 
this limit in Monte Carlo simulations, one usually chooses $L_0 \ll L_1, L_2$ and
applies periodic boundary conditions in the $1$ and $2$ directions.
Note that we shall simulate lattices with $L_1=L_2=L$ throughout. 
As we shall see below, in order to compute surface critical exponents,
the condition $L_0 \ll L_1, L_2$ is not mandatory. Our estimates 
for the surface critical exponents are actually obtained from 
simulations of lattices with $L_0=L_1=L_2=L$.

The reduced Hamiltonian of the Blume-Capel model with film geometry is  
\begin{eqnarray}
\label{Isingaction2}
H &=& - \beta \sum_{<xy>}  s_x s_y + D \sum_x s_x^2  -h \sum_x s_x 
 - \beta_1 \sum_{<xy>,x_0=y_0=1}  s_x s_y 
- \beta_2 \sum_{<xy>,x_0=y_0=L_0}  s_x s_y  \nonumber \\
 &+& D_1 \sum_{x,x_0=1} s_x^2 + D_2 \sum_{x,x_0=L_0} s_x^2 
\;-\; h_1 \sum_{x,x_0=1} s_x
\;-\; h_2 \sum_{x,x_0=L_0} s_x \;\; , 
\end{eqnarray}
where we have put the surfaces at $x_0=1$ and $x_0=L_0$.
In our convention $<xy>$ runs over 
all pairs of nearest neighbor sites with fluctuating spins. Note 
that here the sites $(1,x_1,x_2)$ and $(L_0,x_1,x_2)$ are not nearest 
neighbors as it would be the case for periodic boundary conditions.
For each bulk term there is a corresponding surface enhancement 
term. In our simulations and the analysis of the data we have used
the surface couplings
\begin{equation}
 \bar{\beta}_1 = \beta_1 + \beta  \;\; \mbox{and}  \;\; 
 \bar{\beta}_2 = \beta_2 + \beta
\end{equation}
as parameters instead of the excess surface couplings $\beta_1$ and $\beta_2$.

Note that as long as the bulk transition and the line of surface
transitions remain continuous, the qualitative 
features of the phase diagram that we have discussed in the introduction
should remain unchanged.  Since the values of $D$, $D+D_1$ and $D+D_2$
that we shall consider are much smaller than 
$D_{tri}= 2.0313(4)$ \cite{DeBl04} and 
$D_{tri,2D} =1.966(2)$ \cite{SiCaPl06} of the two-dimensional
Blume-Capel model on the square lattice, this should be the case
in our study. 

\subsection{Observables}
Renormalization group invariant quantities are very useful
to locate critical or multicritical points. We study the 
Binder cumulant 
\begin{equation}
U_4 = \frac{\langle m^4 \rangle} {\langle m^{2} \rangle^2} \;\;,
\end{equation}
where $m=\sum_x s_x$. 
The second moment correlation length is given by
\begin{equation}
\label{xihigh}
\xi_{2nd}  = \sqrt{\frac{\chi/F-1}{4 \sin^2 \pi/L}} \;\;,
\end{equation}
where
\begin{equation}
F =  \frac{1}{L_0 L^2} \left \langle
\Big|\sum_x \exp\left(i \frac{2 \pi x_{1,2}}{L_{1,2}} \right)
        s_x \Big|^2
\right \rangle
\end{equation}
is the Fourier transform of the correlation function at the lowest non-zero
momentum in 1 or 2 direction and $\chi  = \langle m^{2} \rangle/(L_0 L^2)$
is the magnetic susceptibility. 
Since in our simulations $L_1=L_2$, the expectation value of $F$ is identical 
for the 1 and the 2 direction. In order to reduce the statistical error, we have 
measured $F$ for both directions and have averaged the results. The ratio
$\xi_{2nd}/L$ is renormalization group invariant.  The third renormalization group 
invariant quantity that we consider is the 
ratio $Z_a/Z_p$ of partition functions, where $Z_a$ is the partition function
of a system with anti-periodic boundary conditions in  1 direction and
periodic ones in 2 direction or vice versa, while in the case of $Z_p$ periodic
boundary conditions are imposed in 1 and 2 direction. Also here, since $L_1=L_2$
in our simulations, we determine  $Z_a/Z_p$ for both  choices
and average the results. The ratio $Z_a/Z_p$ of partition functions
can be efficiently evaluated using the boundary flip algorithm \cite{BF}.
Here we use a modified  version of the boundary flip algorithm as
discussed in appendix A 2 of ref. \cite{ourXY}.
In the following we shall refer to the renormalization group invariant
quantities $U_4$, $Z_a/Z_p$ and $\xi_{2nd}/L$ using the symbol $R$.
For vanishing symmetry breaking fields $h=h_1=h_2=0$, $D_1=D_2$ and $\beta_1=\beta_2$,
a renormalization group invariant quantity behaves as
\begin{equation}
\label{Rscaling}
 R(L_0,L,\beta,\beta_1) =   
           q(L/L_0,t [L_0/\xi_0]^{y_t},t_1 [L_0/\xi_{1,0}]^{y_{t_1}})  \;\;,
\end{equation}
where $t=\beta_c-\beta$, $t_1=\beta_{1,sp}-\beta_1$, $\xi_0$ is 
the amplitude of the correlation length in the high temperature phase
and the normalisation factor $\xi_{1,0}$ is still undetermined. In section~(\ref{MC1})
we have fixed the ratio $L/L_0=1$ and have set the bulk coupling to its 
critical value: $\beta=\beta_c$.  This allows
us to determine the location of the special point $\beta_{1,sp}$ by using 
the standard crossing method, where $R$ is considered as function of $\beta_1$. 
The behavior of the slope of the renormalization group invariant quantities
allows us to determine the surface RG-exponent $y_{t_1}$:
Taking the derivative with respect to $\beta_1$  we get
\begin{equation}
\label{slopescaling}
\frac{\partial R(L_0,L,\beta,\beta_1)}{\partial \beta_1} =
      Q(L/L_0,t [L_0/\xi_0]^{y_t},t_1 [L_0/\xi_{1,0}]^{y_{t_1}}) \;
      [L_0/\xi_{1,0}]^{y_{t_1}}
\end{equation}
where $Q$ is minus the partial derivative of $q$ with respect to 
its third argument.

Finally we define the surface susceptibilities for a vanishing 
surface magnetisation $\langle m_1 \rangle$:
\begin{equation}
 \chi_{11} = \frac{\partial \langle m_1 \rangle }{ \partial h_1}= L^2 \langle m_1^2  \rangle
\end{equation}
where
\begin{equation}
 m_1 = \frac{1}{L^2} \sum_{x_1,x_2} s_{(1,x_1,x_2)}
\end{equation}
and
\begin{equation}
 \chi_{12} = \frac{\partial \langle m_1 \rangle }{ \partial h_2}= L^2 \langle m_1 m_2  \rangle
\end{equation}
where
\begin{equation}
 m_2 = \frac{1}{L^2} \sum_{x_1,x_2} s_{(L_0,x_1,x_2)} \;\;.
\end{equation}
The finite size scaling behavior of these quantities can be inferred from the 
singular part of the reduced free energy per area of the film. Its scaling form is
\begin{equation}
\label{fsscaling}
 f_s(...,...,...,h_1,h_2) = L_0^{-d+1}
 g(L/L_0,...,...,h_1 [L_0/l_{h_1}]^{y_{h_1}},h_2 [L_0/l_{h_2}]^{y_{h_2}}) \;\;,
\end{equation}
where the constants $l_{h_1}$  and $l_{h_2}$ remain undetermined here. $y_{h_1}$
and $y_{h_2}$ are the RG-exponents related to the surface fields.
The susceptibilities defined above can be expressed as second derivatives of $f_s$  
with respect to the surface fields. Taking these derivatives on the right hand 
side of eq.~(\ref{fsscaling}) we arrive at 
\begin{equation}
\label{chiscaling1}
 \chi_{11} =  \frac{\partial^2 f_s}{\partial h_1^2} = 
  c L_0^{-d+1} [L_0/l_{h_1}]^{2 y_{h_1}} 
\end{equation}
and
\begin{equation}
\label{chiscaling2}
\chi_{12} = \frac{\partial^2 f_s}{\partial h_1 \partial h_2} 
=c L_0^{-d+1} [L_0/l_{h_1}]^{y_{h_1}} [L_0/l_{h_2}]^{y_{h_2}} \;\;.
\end{equation}
In our study below, section \ref{SurfaceSus}, $y_{h_1}=y_{h_2}$ and $l_{h_1}=l_{h_2}$,
since we have chosen $\beta_1=\beta_2$, $D_1=D_2$ and $h_1=h_2=0$.

\subsection{Corrections to scaling}
\label{corrections}
Finite size scaling laws such as 
eqs.~(\ref{Rscaling},\ref{slopescaling},\ref{chiscaling1},\ref{chiscaling2}) 
are affected by
corrections. These are caused by irrelevant scaling fields, the analytic 
background in the reduced free energy per area and the fact that e.g. in 
eq.~(\ref{Rscaling}) the arguments of the scaling function $q$ should 
be actually analytic functions of $t$ and $t_1$ that we have linearized 
here. 

The leading bulk correction is $\propto L_0^{-\omega}$, where our most recent 
estimate $\omega=0.832(6)$ obtained from Monte Carlo simulations of the 
Blume-Capel and the Ising model \cite{mycritical} is slightly larger than that 
obtained from field theoretical methods, e.g. $\omega=0.799(11)$ by using 
perturbation theory in three dimensions fixed and $\omega=0.814(14)$ 
by using the $\epsilon$-expansion \cite{GuZi98}. There are also corrections
$\propto L_0^{-n \omega}$, where $n=2,3,...$\phantom{.}. 
In the case of improved models these 
are highly suppressed and therefore we shall ignore them in the analysis of our 
data below.
Following \cite{NewmanRiedel} subleading corrections are characterized by 
$\omega_2 =1.67(11)$. There is no reason to assume that the  amplitude 
of the subleading correction vanishes in the case of the improved Blume-Capel 
model. Likely it is of similar size as in the spin-1/2 Ising model.
Furthermore there are well established corrections with $\omega_i \approx 2$,
for example related to the breaking of the spatial rotational invariance by the 
simple cubic lattice \cite{CPRV-99}.

In the presence of surfaces there are also corrections caused by irrelevant surface 
fields. One expects that the leading ones are $\propto L_0^{-1}$ 
\cite{BinderS,Diehl86,Diehl97}.

In general we expect that for finite $L_0$ finite size scaling laws such as 
eqs.~(\ref{slopescaling},\ref{chiscaling1},\ref{chiscaling2}) can be written 
in the form  of a Wegner-expansion \cite{wegner}
\begin{equation}
 A(L_0) = c L_0^{x} [ 1 + a
L_0^{-\omega} + \sum_{i=2} a_i L_0^{-\omega_i} ] \;\;
\label{general}
\end{equation}
with an infinite number of correction terms.  Fitting data obtained from
Monte Carlo simulations, only a very limited number of terms can be taken 
into account. This unavoidable truncation of the Wegner-expansion leads to 
a systematic error of the estimate of e.g. the exponent $x$ that is often larger 
than the statistical one.

In the present work, we shall use ansaetze that include either no correction, 
a correction $\propto  L_0^{-1}$ or corrections $\propto  L_0^{-1}$ and 
$\propto  L_0^{-2}$. In the case of the surface susceptibilities we shall also
take into account a term for the analytic background. The systematic errors of our
results are then estimated from the variation of the results obtained by using 
these different ansaetze.

\section{Simulations of $L^3$ lattices at the special point}
\label{MC1}
In this part of our study we consider three different models. In all cases 
we chose $L_0=L$, $\bar{\beta}_1=\bar{\beta}_2$, $D_1=D_2$ 
and $h_1=h_2=0$. We have 
simulated the Blume-Capel model at $D=0.655$ with two different choices of $D_1$: 
The choice $D_1=0$ is called BC1 model and $D_1 \rightarrow - \infty$ 
is called BC2 model in the following. Note that in the case
of the BC2 model, the spins at the surfaces can take only the values
$-1$ or $1$. In addition, we have simulated the spin-1/2 Ising model.
We have simulated at the best estimates of the bulk critical point; i.e. 
$\beta_c=0.387721735$ for the Blume Capel model at $D=0.655$ and 
$\beta_c=0.2216546$ for the spin-1/2 Ising model. The numerical 
uncertainty of these numbers is negligible in the present study. 

For each measurement of the observables we performed the following 
sequence of Monte Carlo updates: First one sweep with a local update.
In the case of the Ising model we use a local Metropolis and for the 
Blume-Capel model a local heat-bath update. One sweep means that we run
through the lattice once in type-writer fashion. Then we performed a
certain number of single-cluster updates \cite{Wolff} followed 
by two wall-cluster updates \cite{wall}; one for each of the directions with 
periodic boundary conditions. In all our simulations we have used the 
Mersenne twister algorithm \cite{twister} as pseudo-random number generator.

In the case of the Blume-Capel models BC1 and BC2, no previous 
estimate of the surface coupling $\bar{\beta}_{1,sp}$ was available.
Therefore, we have successively improved our estimate of $\bar{\beta}_{1,sp}$ 
with increasing lattice size $L$. 
In order to obtain the observables as a function of $\bar{\beta}_1$ 
in the neighborhood of the simulation point, we have computed the
coefficients of the Taylor-expansion of the observables in $\bar{\beta}_1$ 
around the simulation point up to third order.
In table \ref{statQ} we summarize the lattice sizes that we have simulated 
at and the statistics of these simulations.  In total we have spent about 12 
years of CPU time on a single
core of a  Quad-Core AMD Opteron(tm) Processor 2378 running at 2.4 GHz.

\begin{table}
\caption{\sl \label{statQ} Number of measurements divided by $10^6$. For each 
measurement we perform one heat-bath sweep, a certain number of single cluster 
and two wall-cluster updates.
}
\begin{center}
\begin{tabular}{rrcc}
\hline
\mc{1}{c}{$L$} &\mc{1}{c}{BC1}& \mc{1}{c}{BC2} &\mc{1}{c}{Ising} \\ 
\hline
   8   & 1000  &  1000    &   1000 \\
  12   & 1000  &  1000    &   1000 \\
  16   & 1000  &  1000    &   1000 \\
  24   & 1000  &  1178    &    877 \\
  32   & 1020  &  1042    &    718 \\
  48   &  861  &   705    &    305 \\
  64   &  593  &   482    &    224 \\
  96   &  301  &   200    &    132 \\
 128   &  202  &   140    &    116 \\
\hline
\end{tabular}
\end{center}
\end{table}

In our simulations we have measured the renormalization group invariant
quantities  $Z_a/Z_p$, 
$U_4$ and $\xi_{2nd}/L$. Analysing the data, it turned out that corrections 
to scaling are considerably larger for $\xi_{2nd}/L$  than for $Z_a/Z_p$ and 
$U_4$. Therefore we restrict the following discussion to $Z_a/Z_p$ and $U_4$.

In a first step of the analysis we have computed the surface
coupling $\bar{\beta}_{1,sp}$ of the special point and the fixed point values 
of $Z_a/Z_p$ and $U_4$. To this end we have fitted the data with the ansatz
\begin{equation}
\label{fitR1}
 R(\bar{\beta}_{1,sp} , L) = R^*  
\end{equation}
where $\bar{\beta}_{1,sp}$ and $R^*$ are the free parameters of the fit and
\begin{equation}
 R(\bar{\beta}_{1,sp} , L) = R(\bar{\beta}_{1,s} , L) + 
c_1 (\bar{\beta}_{1,sp}-\bar{\beta}_{1,s}) 
 + \frac{c_2}{2!} (\bar{\beta}_{1,sp}-\bar{\beta}_{1,s})^2
+ \frac{c_3}{3!}  (\bar{\beta}_{1,sp}-\bar{\beta}_{1,s})^3
\end{equation}
where $R(\bar{\beta}_{1,s}, L)$, $c_1$, $c_2$ and $c_3$ are obtained from the 
simulation at $\bar{\beta}_{1,s}$.  
To check for the possible effect of corrections 
to scaling we have also used the ansaetze 
\begin{equation}
\label{fitR2}
R(\bar{\beta}_{1,sp} , L) = R^*  + c L^{-1} 
\end{equation}
and
\begin{equation}
\label{fitR3}
R(\bar{\beta}_{1,sp} , L) = R^*  + c L^{-1} + d L^{-2}  \;\;.
\end{equation}
First we have analyzed the three models separately. Let us first look 
at the results for the BC1 model and the ratio $Z_a/Z_p$. A selection 
of  results is given in table \ref{Fit1}. In these fits we have taken 
into account the data for all lattice sizes $L \ge L_{min}$.
Fits with an acceptable 
$\chi^2$/DOF are obtained starting from $L_{min}=48$, $24$ and $16$ for
the ansaetze~(\ref{fitR1}), (\ref{fitR2}) and (\ref{fitR3}), respectively.
Note that the differences of the results for $\bar{\beta}_{1,sp}$ 
and $(Z_a/Z_p)^*$
for different ansaetze with an acceptable $\chi^2$/DOF are larger than 
the statistical errors.  Next we have fitted the Binder cumulant with 
the ansaetze~(\ref{fitR1}), (\ref{fitR2}) and (\ref{fitR3}).   In the 
case of the ansatz~(\ref{fitR1})  we find $\chi^2$/DOF$=1.32/4$ already 
for $L_{min}=24$ with $\bar{\beta}_{1,sp}=0.5491393(32)$ and 
$U_4^*=1.52338(7)$. 
Fitting the data for $L_{min}=8$ with the ansaetze~(\ref{fitR2}) and 
(\ref{fitR3}) we get $\chi^2$/DOF$=8.69/6$  and $8.61/5$, respectively.
The results for the fit parameters are $\bar{\beta}_{1,sp}=0.5491470(37)$, 
$U_4=1.52303(10)$ and $\bar{\beta}_{1,sp}=0.5491455(63)$, $U_4=1.52309(23)$,
for the ansaetze~(\ref{fitR2}) and (\ref{fitR3}), 
respectively. Notice that the results for $\bar{\beta}_{1,sp}$ are compatible
with those obtained from the analysis of $Z_a/Z_p$. 

Next we have analyzed the data for the BC2 model. In the case of the 
ansatz~(\ref{fitR1}) larger $L_{min}$ are need than for the BC1 model
to get an acceptable $\chi^2$/DOF.  Fitting the data with the 
ansaetze~(\ref{fitR2}) and (\ref{fitR3}) we see that the correction 
amplitude $c$ is clearly larger for the BC2 than for the BC1 model.

\begin{table}
\caption{\sl \label{Fit1} Fitting the ratio of partition functions $Z_a/Z_p$
for the BC1 model with the ansaetze~(\ref{fitR1}), (\ref{fitR2}) or 
(\ref{fitR3}). 
}
\begin{center}
\begin{tabular}{cccllrc}
\hline
\mc{1}{c}{Ansatz} &\mc{1}{c}{$L_{min}$}&\mc{1}{c}{$\bar{\beta}_{1,sp,BC1}$}&
\mc{1}{c}{$(Z_a/Z_p)^*$} & \mc{1}{c}{$c$} & \mc{1}{c}{$d$} & 
\mc{1}{c}{$\chi^2/$DOF}  \\
\hline
\ref{fitR1}   &  32  &  0.5491558(23) & 0.31318(5) &     &    & 14.2/3 \\
\ref{fitR1}   &  48  &  0.5491482(32) & 0.31341(8) &     &    &  2.32/2 \\
\ref{fitR2}   &  16  &  0.5491430(33) & 0.31374(10)&0.0131(14)&  & 6.08/4 \\
\ref{fitR2}   &  24  &  0.5491350(47) & 0.31406(17)&0.0194(31)&  & 0.69/3 \\
\ref{fitR3}&12& 0.5491423(49) & 0.31375(19)&0.0121(47)&0.023(34)&
8.87/4 \\
\ref{fitR3}&16& 0.5491280(69) & 0.31445(32)&0.035(10) & --0.216(94)&0.51/3\\
\hline
\end{tabular}
\end{center}
\end{table}

Next we performed joint fits for the BC1 and the BC2 model using 
the ansaetze~(\ref{fitR2}) and (\ref{fitR3}). 
In these fits we impose that, following universality, $(Z_a/Z_p)^*$ and 
$U_4^*$ are the same for both models. A selection of results is given 
in table \ref{FitJ}. We see that the correction
amplitude $c_2$ of the BC2 model is  much larger than $c_1$ of the BC1 model.  
We performed 
similar fits for the Binder cumulant $U_4$.  Also here we observe that
the correction $\propto L^{-1}$ has a much larger amplitude for the 
BC2 model than for the BC1 model. The results for the surface 
couplings at the special point $\bar{\beta}_{1,sp,BC1}$ and 
$\bar{\beta}_{1,sp,BC2}$ obtained from the analysis
of $U_4$ and $Z_a/Z_p$ are compatible among  each other. 

\begin{table}
\caption{\sl \label{FitJ} Fitting the ratio of partition functions $Z_a/Z_p$
for the BC1 and BC2 model with the ansaetze~(\ref{fitR2}) or
(\ref{fitR3}). In the case of ansatz~(\ref{fitR3}) we do not report $d_1$ 
and  $d_2$ to keep the table readable.
}
\begin{center}
\begin{tabular}{crccllrc}
\hline
\mc{1}{c}{Ansatz} &\mc{1}{c}{$L$}&
\mc{1}{c}{$\bar{\beta}_{1,sp,BC1}$}&  \mc{1}{c}{$\bar{\beta}_{1,sp,BC2}$}& 
\mc{1}{c}{$(Z_a/Z_p)^*$} & \mc{1}{c}{$c_1$} & \mc{1}{c}{$c_2$} &
\mc{1}{c}{$\chi^2/$DOF}  \\
\hline
\ref{fitR2} & 12 & 0.5491418(20) & 0.2940123(19) & 0.31380(5) & 
                0.0144(6)&--0.0782(6)\phantom{0} &19.78/11 \\
\ref{fitR2} & 16 & 0.5491461(24) & 0.2940165(23) & 0.31364(7) &
                                            0.0118(10)&--0.0806(10)& 9.36/9 \\
\ref{fitR2} & 24 & 0.5491416(35) & 0.2940124(33) &  0.31382(13) &
                                            0.0152(23) & --0.0771(22)&6.20/7 \\

\ref{fitR3} &  8 & 0.5491493(27) & 0.2940202(26) & 0.31346(9)&
                                           0.0045(17) & --0.0863(16)&15.44/11\\
\ref{fitR3} & 12 & 0.5491481(28) & 0.2940189(36) & 0.31352(15)&
    0.0064(38) & --0.0850(37)&15.13/9\phantom{0} \\ 
\ref{fitR3} & 16 & 0.5491364(52) & 0.2940086(49) & 0.31407(23)& 0.0242(70) &
--0.0664(68) & 4.70/7\\
\hline
\end{tabular}
\end{center}
\end{table}

\begin{table}
\caption{\sl \label{FitJJ} Same as table \ref{FitJ} but for the 
Binder cumulant $U_4$ instead of $Z_a/Z_p$. 
}
\begin{center}
\begin{tabular}{crcclllc}
\hline
\mc{1}{c}{Ansatz} &\mc{1}{c}{$L$}&
\mc{1}{c}{$\bar{\beta}_{1,sp,BC1}$}&  \mc{1}{c}{$\bar{\beta}_{1,sp,BC2}$}&
\mc{1}{c}{$U_4^*$} & \mc{1}{c}{$c_1$} & \mc{1}{c}{$c_2$} &
\mc{1}{c}{$\chi^2/$DOF}  \\
\hline
\ref{fitR2} & 16 & 0.5491396(42) & 0.2940168(40) & 1.52327(13) &
--0.005(2)  & 0.347(2) & 16.75/9 \\
\ref{fitR2} & 24 & 0.5491377(61) & 0.2940104(57) & 1.52345(23) &
\phantom{--}0.002(4)  & 0.348(4) & 1.95/7 \\
\ref{fitR3} & 8  & 0.5491520(45) & 0.2940194(41) & 1.52285(15) &
--0.011(3)  & 0.323(3) & 13.58/11 \\
\ref{fitR3} &12  & 0.5491517(66) & 0.2940202(62) & 1.52284(27) &
--0.012(7)  &0.324(7)  &  13.00/9\phantom{0} \\
\ref{fitR3} &16  & 0.5491342(82) & 0.2940047(80) & 1.52372(38) &
\phantom{--}0.017(12) & 0.353(11) & 5.84/7 \\
\hline
\end{tabular}
\end{center}
\end{table}

As our final result we quote
\begin{equation}
\label{fixedpoint}
 (Z_a/Z_p)^* =   0.3138(5)  \;\;,\;\; U_4^* = 1.5234(10)
\end{equation}
and
\begin{equation}
\label{finalsp}
 \bar{\beta}_{1,sp,BC1} = 0.54914(2) \;, \;\; \bar{\beta}_{1,sp,BC2}  = 
 0.29401(2)
\end{equation}
where the error bars are chosen such that the results of all fits given in 
table \ref{FitJ} and \ref{FitJJ} are covered. 

Finally we have fitted our data for the spin-1/2 Ising model using 
the ansatz~(\ref{fitR2}).
A selection of results is given in table \ref{FitIs}. For $L_{min}=16$ and 
$24$ we find acceptable values for $\chi^2/$DOF. Nevertheless the result
for $(Z_a/Z_p)^*$ is not compatible with that obtained from the fits for the 
BC1 and BC2 models. This could be explained by the fact that 
corrections $\propto L^{-\omega}$ with $\omega=0.832(6)$ \cite{mycritical}
are not explicitly taken into account.

\begin{table}
\caption{\sl \label{FitIs} Fitting the ratio of partition functions $Z_a/Z_p$
for the spin-1/2 Ising model with the ansatz~(\ref{fitR2}).
}
\begin{center}
\begin{tabular}{ccllc}
\hline
\mc{1}{c}{$L$}&\mc{1}{c}{$\bar{\beta}_{1,sp,I}$}&
\mc{1}{c}{$(Z_a/Z_p)^*$} & \mc{1}{c}{$c$} & \mc{1}{c}{$\chi^2/$DOF}  \\
\hline

12  & 0.3330388(15) & 0.31221(5) & 0.00219(5)  & 9.03/5 \\
16  & 0.3330365(19) & 0.31232(7) & 0.00236(9)  & 4.34/4 \\
24  & 0.3330338(28) & 0.31246(13)& 0.00262(23) & 2.77/3 \\
\hline
\end{tabular}
\end{center}
\end{table}

Finally we performed a number of different fits, where we take the 
results~(\ref{fixedpoint}) for $(Z_a/Z_p)^*$ and $U_4^*$ as input. In these
fits we also include explicitly corrections $\propto L^{-\omega}$. Taking 
into account these different fits we arrive at the final estimate
\begin{equation}
 \bar{\beta}_{1,sp,I} = 0.33302(2)
\end{equation}
for the spin-1/2 Ising model. In table \ref{betasc} we compare our 
result with those given in the literature.
\begin{table}
\caption{\sl \label{betasc} Comparison of our result for the ratio
$\bar{\beta}_{1,sp,I}/\beta_{c}$ with those given in the literature.
}
\begin{center}
\begin{tabular}{ccl}
\hline
 Ref.             & year & $\bar{\beta}_{1,sp,I}/\beta_{c} $\\
\hline
\cite{BiLa84}     & 1984 &  1.50(3) \\
\cite{LaBi90}     & 1990 &  1.52(2) \\
\cite{RuDuWaWu93} & 1993 &  1.5004(20)  \\ 
\cite{DeBlNi05}   & 2005 &  1.50208(5) \\
  Here            & 2011 &  1.50243(9) \\
\hline
\end{tabular}
\end{center}
\end{table}
Our result is compatible within error bars with all others 
except for the one of ref. \cite{DeBlNi05}, where  we find that 
the difference is 2.5 times larger than the combined error. 

\subsection{The RG-exponent  $y_{t_1}$}
We have determined the  critical exponent  $y_{t_1}$ from the slope 
of a renormalization group invariant quantity $R_1$ at a fixed value 
$R_{2,f}$ of a second renormalization group invariant quantity $R_2$ 
\begin{equation}
\bar{S} := 
\left . \frac{\partial R_1}{\partial \beta_1} \right |_{R_2 = R_{2,f}}
 \simeq c L^{y_{t_1}} \;\;, \\
\end{equation}
where in our case $R_1$ is either the Binder cumulant $U_4$ or the ratio 
of partition functions $Z_a/Z_p$ and we fix  $(Z_a/Z_p)_f=0.3138$.
Fitting our data we have used the ansaetze
\begin{equation}
\label{slope1}
 \bar{S} = c L^{y_{t_1}}  \;\;,
\end{equation}
\begin{equation}
\label{slope2}
 \bar{S} = c L^{y_{t_1}}  (1 + d L^{-1}) \;\;,
\end{equation}
and
\begin{equation}
\label{slope3}
 \bar{S} = c L^{y_{t_1}}  (1 + d L^{-1}+ e L^{-2}) \;\;.
\end{equation}
In  table \ref{FitSlope} the results of
fits for the slope of $Z_a/Z_p$ at $Z_a/Z_p=0.3138$ for the BC1 model
are given.
We also performed joint fits of  the BC1 and the BC2 model.
In table \ref{FitSlope2} we report results obtained from 
the slope of $Z_a/Z_p$ at $Z_a/Z_p=0.3138$. 
Also in the case of the slopes we observe that the amplitude 
of $L^{-1}$ corrections is much larger for the BC2 than for the BC1 model.
Therefore using ansatz~(\ref{slope1}) no acceptable fit can be obtained.
\begin{table}
\caption{\sl \label{FitSlope} 
Fitting the slope of $Z_a/Z_p$ at $Z_a/Z_p=0.3138$ for the BC1 model.
}
\begin{center}
\begin{tabular}{cccllc}
\hline
\mc{1}{c}{Ansatz} &\mc{1}{c}{$L_{min}$}&\mc{1}{c}{$y_{t_1}$}&
\mc{1}{c}{$d$} & \mc{1}{c}{$e$} & \mc{1}{c}{$\chi^2/$DOF}  \\
\hline
\ref{slope1} & 32&  0.71909(21) &    &  & 4.26/3 \\
\ref{slope1} & 48&  0.71855(36) &    &  & 0.89/2 \\
\ref{slope2} & 12&  0.71583(28) & --0.148(6)  &   & 20.91/5 \\
\ref{slope2} & 16&  0.71700(38) & --0.108(11) &   & 1.57/4 \\
\ref{slope3} & \phantom{0}8 &  0.71795(49) & --0.026(21) &  --0.71(9) & 5.39/5 \\
\ref{slope3} &12 &  0.71872(82) & \phantom{--}0.022(49) & --1.01(30) & 3.75/4\\
\hline
\end{tabular}
\end{center}
\end{table}

\begin{table}
\caption{\sl \label{FitSlope2}
Fitting the slope of $Z_a/Z_p$ at $Z_a/Z_p=0.3138$ for the BC1 and the BC2 
model jointly.
}
\begin{center}
\begin{tabular}{cccccccc}
\hline
Ansatz &$L_{min}$& $y_{t_1}$ &$d_1$ & $d_2$ &$e_1$ & $e_2$ &{$\chi^2/$DOF}\\ 
\hline
 \ref{slope2}& 16 & 0.71694(29) & --0.109(9)\phantom{0} &  0.354(9)&  &  & 7.18/9 \\
 \ref{slope2}& 24 & 0.71758(50) & --0.084(21)&  0.386(21)& &  &  3.76/7 \\
 \ref{slope3}& \phantom{0}8 & 0.71792(29) & --0.027(11)&  0.436(13)& --0.71(4)\phantom{0} & --0.69(5)\phantom{0}  &  10.87/11\\
 \ref{slope3}& 12 & 0.71898(60) &\phantom{--}0.037(32)&0.505(32)& --1.09(19)  & --1.12(19) &4.86/9\\
\hline
\end{tabular}
\end{center}
\end{table}
Based on these fits we arrive at the final estimate
\begin{equation}
\label{final_yst}
 y_{t_1} = 0.718(2) \;\;.
\end{equation}
The central value and the error bar are chosen such that all the results 
of the fits  reported in tables \ref{FitSlope} and \ref{FitSlope2} are 
covered. Our final estimate is also consistent with the results 
obtained from the slope of the Binder cumulant $U_4$. We have also fitted 
our data for the Ising model with the 
ansaetze~(\ref{slope1},\ref{slope2},\ref{slope3}). The results
obtained from the slope of $Z_a/Z_p$ for $y_{t_1}$ are fully consistent 
with the estimate~(\ref{final_yst}), while those from the slope of  $U_4$
are a bit larger. We conclude that leading bulk corrections 
$\propto L^{-\omega}$ have only a small numerical effect in the case of the 
Ising model and can therefore be savely ignored in the case of the 
Blume-Capel model at $D=0.655$, where leading corrections to scaling 
are suppressed at least by a factor of $30$ compared with the Ising model.

\subsection{Surface susceptibilities and the RG-exponent $y_{h_1}$}
\label{SurfaceSus}
Finally we have determined the exponent  $y_{h_1}$ from the finite size 
scaling behavior of the surface susceptibilities $\chi_{11}$ and 
$\chi_{12}$. To this end we have
computed
\begin{equation}
\bar{\chi} := \chi |_{R=R_f}  \propto L^{ 2y_{h_1} -2}
\end{equation}
where we use $(Z_a/Z_p)_f=0.3138$. We have fitted these quantities 
with the ansaetze
\begin{equation} 
\label{chi0}
 \bar{\chi} = c L^x \;\;
\end{equation}
where $x=2y_{h_1} -2$,
\begin{equation}
\label{chi1}
 \bar{\chi} = c L^x (1 + d L^{-1}) \;,
\end{equation}
\begin{equation}
\label{chi2}
 \bar{\chi} = c L^x (1 + d L^{-1}) + b 
\end{equation}
where $b$ is an analytic background and 
\begin{equation}
\label{chi3}
 \bar{\chi} = c L^x (1 + d L^{-1} + e L^{-2} ) + b \;\;.
\end{equation}

\begin{table}
\caption{\sl \label{Fitchi}
Fitting the surface susceptibilities $\chi_{11}$ and $\chi_{12}$ at $Z_a/Z_p=0.3138$.
Both the data of the BC1 and the BC2 model are taken into account.
Results for the parameters $b$ and $e$ are not reported to keep the 
table readable.
}
\begin{center}
\begin{tabular}{ccccccc}
\hline
Quantity & Ansatz &$L_{min}$& $ x $ &$d_1$ & $d_2$ &{$\chi^2/$DOF}\\
\hline
$\bar{\chi}_{11}$&\ref{chi1}& 24 &1.2930(3)& --0.09(1)  & 1.13(1)   & 10.37/7 \\
$\bar{\chi}_{11}$&\ref{chi1}& 32 &1.2932(4)& --0.09(2)  & 1.15(2)   & 4.48/5 \\
$\bar{\chi}_{11}$&\ref{chi2}&16 & 1.2935(4)& --0.08(5)  & 1.39(6)   & 6.87/7 \\
$\bar{\chi}_{11}$&\ref{chi3}&8  & 1.2933(6)& --0.19(15) & 1.12(15)  & 8.56/9 \\

$\bar{\chi}_{12}$&\ref{chi1}&24 & 1.2921(2)& --0.36(1)  & 0.02(1)   & 4.01/7 \\
$\bar{\chi}_{12}$&\ref{chi1}&32 & 1.2922(3)& --0.35(1)  & 0.03(1)   & 2.96/5 \\
$\bar{\chi}_{12}$&\ref{chi2}&16 & 1.2932(4)& --0.19(6)  & 0.35(6)   & 10.28/7 \\
$\bar{\chi}_{12}$&\ref{chi2}&24 & 1.2927(5)& --0.25(12) & 0.16(13)  & 3.10/5 \\
$\bar{\chi}_{12}$&\ref{chi3}&8  & 1.2935(2)& --0.13(2)  & 0.32(5)   & 14.61/9 \\
$\bar{\chi}_{12}$&\ref{chi3}&12 & 1.2929(3)& --0.30(4)  & 0.02(9)   & 7.83/7 \\
$\bar{\chi}_{12}$&\ref{chi3}&16 & 1.2928(6)& --0.14(17) & --0.06(25)& 3.19/5 \\
\hline
\end{tabular}
\end{center}
\end{table}

It turns out that using the ansatz~(\ref{chi0}) for none 
of the models studied an acceptable fit can be obtained.
In table \ref{Fitchi} results of joint fits of the BC1 and the BC2 model
using the ansaetze~(\ref{chi1},\ref{chi2},\ref{chi3}) are reported.
The variation of the results for the parameter $x$ with the different ansaetze 
is of similar size as the statistical errors. The results obtained 
from $\chi_{11}$ are slightly larger than those from $\chi_{12}$. Our final
estimate $x=1.2929(10)$ covers all results reported in table \ref{Fitchi}.
Fitting the data obtained for the Ising model we get results for the 
parameter $x$ that are essentially consistent with that obtained for 
the Blume-Capel models BC1 and BC2. Therefore residual leading bulk 
corrections in the case of the  Blume-Capel model at $D=0.655$ should
not affect our final estimate of $x$. As our final estimate for the 
RG-exponent of the external surface field we quote
\begin{equation}
\label{yshfinal}
y_{h_1} = 1.6465(6) \;\;.
\end{equation}

\section{Numerical results for the surface transition}
In the neighborhood of the special point, the transition line behaves 
as \cite{BinderS,Diehl86,Diehl97}
\begin{equation}
\label{scalingt}
 |t_{1,c}| = c |t|^{\Phi}  \;\;,
\end{equation}
where $\Phi=\nu y_{t_1}$  is the cross-over exponent. In the present section
we shall compute the line of surface transitions numerically and check how
well it is described by eq.~(\ref{scalingt}). To this end,
we have only simulated the BC1 model, i.e. the Blume-Capel model at $D=0.655$ 
and $D_1=D_2=0$, since it has the smallest scaling corrections among the three
models discussed in the preceding section. 
For given values of the bulk coupling $\beta<\beta_c$ we have 
determined the critical surface coupling  $\bar{\beta}_{1,c}$. To this end
we have simulated films with an excess surface coupling $\beta_1>0$ at one 
surface and free boundary conditions, i.e. $\beta_2=0$, at the other.
For $L_0 \gg \xi$, where $\xi$ is the correlation length of the bulk, the 
film should provide a good approximation
of the semi-infinite system. We have determined $\bar{\beta}_{1,c}$ by requiring 
that the ratio of partition functions $Z_a/Z_p$ assumes the fixed point 
value of a square system with periodic boundary conditions in the universality
class of the two-dimensional Ising model \cite{CaHa95}
\begin{equation}
\label{2Dfix}
(Z_a/Z_p)^*=0.372884880824589... \;.
\end{equation}
In the case of $Z_a/Z_p$ corrections to the fixed point value of the 
two-dimensional Ising universality class vanish $\propto L^{-4}$
\cite{CaHaPeVi02}. Note that in the case of $U_4$ and $\xi_{2nd}/L$ the analytic 
background of the magnetisation enters. Therefore, for these
two quantities corrections vanish $\propto L^{-7/4}$.

Similar to the previous section, we have updated the configurations by 
using a hybrid of the local heat-bath algorithm, the single-cluster
algorithm and the wall-cluster algorithm. 
First we performed a series of simulations with varying values
of $L_0$ and $L$ at $\beta=0.31$, where  $\xi_{2nd}=1.04430(21)$.
Our results are summarized in table \ref{table31}.
We expect that corrections to the limit $L_0,L \rightarrow \infty$ decay 
exponentially with increasing $L_0$ and power-like as $L$ increases.
\begin{table}
\caption{\sl  \label{table31}   Value of $\bar{\beta}_{1,c}$  
obtained by requiring that $Z_a/Z_p=0.372884880824589...$  as a function 
of $L_0$ and $L$. Throughout $\beta=0.31$. 
}
\begin{center}
\begin{tabular}{rrl}
\hline
 \mc{1}{c}{$L_0$} &\mc{1}{c}{$L$}  & \mc{1}{c}{$\bar{\beta}_{1,c}$} \\
\hline
   10 &   8 & 0.6738189(96) \\
   10 &  10 & 0.6739017(78) \\
   10 &  12 & 0.6739022(65) \\
   10 &  20 & 0.6738941(40) \\
   10 & 100 & 0.6738924(95) \\
\hline
    5 & 100 & 0.6739308(83) \\
    6 & 100 & 0.6739055(86) \\ 
    7 & 100 & 0.6738980(85) \\ 
    8 & 100 & 0.6738761(85) \\ 
   12 & 100 & 0.6738845(82) \\ 
\hline 
\end{tabular}
\end{center}
\end{table}
We see that for $L=20$ and $L_0=10$ the corrections
are smaller than the statistical error of our numerical results. Assuming
scaling, we chose for our simulations at larger values of $\beta$ the 
lattice size such that $L_0 > 10 \xi$ and $L = 2 L_0$. The numerical results
of these simulations are summarized in table \ref{surfacetrans}. 
In total these simulations took about 8 month of CPU time on a single core
of a Quad-Core AMD Opteron(tm) Processor 2378 running at 2.4 GHz.

\begin{table}
\caption{\sl  \label{surfacetrans} 
The surface coupling $\bar{\beta}_{1,c}$ at the surface transition 
for various values of the 
bulk coupling $\beta$. In addition we give the second moment 
correlation length $\xi_{2nd}$ of the bulk system, the linear lattice 
sizes $L_0$ and $L$ and the number of measurements (stat).
}
\begin{center}
\begin{tabular}{lrrrrr}
\hline
\mc{1}{c}{$\beta$} & \mc{1}{c}{$\xi_{2nd}$} & 
\mc{1}{c}{$L_0$} & \mc{1}{c}{$L$} & \mc{1}{c}{stat$/10^5$} & 
\mc{1}{c}{$\bar{\beta}_{1,c}$}  \\ 
\hline
0.36  & 2.1272(4)\phantom{0} & 24 &  48 &  100\phantom{.0} &  0.6394276(58) \\
0.378 & 4.1937(7)\phantom{0} & 50 & 100 &  100\phantom{.0} &  0.6093666(31) \\
0.3822& 6.0133(7)\phantom{0} & 70 & 140 &  100\phantom{.0} &  0.5967239(25) \\ 
0.3842& 7.9984(13)&           100 & 200 &   35\phantom{.0} &  0.5883957(31) \\ 
0.3859&12.1358(22)&           150 & 300 &   20.2   &          0.5785822(31) \\ 
0.3865&15.6127(28)&           180 & 360 &   22.5   &          0.5738132(27) \\
0.3869&20.0576(45)&           250 & 500 &    6.9   &          0.5698232(38) \\ 
\hline
\end{tabular}
\end{center}
\end{table}

In order to check the 
prediction~(\ref{scalingt}) we have fitted our data with the 
ansatz 
\begin{equation}
\label{tsansatz}
 \bar{\beta}_{1,c} = \bar{\beta}_{1,sp} 
+ a \left(t+ \sum_{i=2}^m c_i t^i \right)^{\Phi}
\end{equation}
where we have fixed $\Phi=\nu y_{t_1}=0.63002 \times 0.718$. 
Our results are summarized in table \ref{fitS}.  
\begin{table}
\caption{\sl \label{fitS} 
Fitting the critical surface coupling $\bar{\beta}_{1,c}$ with the
ansatz~(\ref{tsansatz}), where $m-1$ gives the number of correction terms that 
are included. For a discussion see the text.
}
\begin{center}
\begin{tabular}{clllllc}
\hline
\mc{1}{c}{$m$} & \mc{1}{c}{$\bar{\beta}_{1,min}$} &
\mc{1}{c}{$\bar{\beta}_{1,sp}$} & \mc{1}{c}{$a$} & \mc{1}{c}{$c_2$} & \mc{1}{c}{$c_3$} &
\mc{1}{c}{$\chi^2/$DOF}  \\
\hline
 3 &  0.36   & 0.549350(9) & 0.51156(18) & --11.27(7) & 114.7(1.9) & 51.6/3 \\
 3 &  0.378  & 0.549222(20)& 0.51490(50) & --13.60(33)& 266(21)  & 0.77/2 \\
 2 &  0.378 &  0.549454(7) & 0.50893(14) & --9.45(4)  & & 154/3 \\
 2 &  0.3822 & 0.549310(14)& 0.51236(32) & --10.99(13)& & 10.1/2 \\
 2 &  0.3842 & 0.549240(26)& 0.51419(66) & --12.15(39)& & 0.00/1 \\
 1 &  0.3842 & 0.549993(7) & 0.49483(12) &           & & 877/2 \\
 1 &  0.3859 & 0.549642(14)& 0.50214(28) &           & & 28.8/1 \\
 1 &  0.3865 & 0.549519(27)& 0.50492(59) &           & &  --   \\
\hline
\end{tabular}
\end{center}
\end{table}
We see that without any correction, even the fit that includes only the 
three largest 
values of $\beta$ that we have simulated has a very large $\chi^2/$DOF and 
the result for $\bar{\beta}_{1,sp}$ is by far too large compared with that obtained
in the previous section. Adding corrections terms, smaller values of 
$\beta$ can be included into the fits. Also the value of $\bar{\beta}_{1,sp}$  gets 
closer to our result obtained  above when adding further correction terms.
However, the values of the correction
amplitudes are large and rapidly increase with increasing order.
We conclude that the behavior of the 
surface transition seems to be consistent with the theoretical 
expectation~(\ref{scalingt}). However an accurate estimate of 
$\bar{\beta}_{1,sp}$
cannot be obtained from the numerical analysis of the surface transition.
Finally we have performed a fit, where we have fixed $\bar{\beta}_{1,sp}$ 
as well as the crossover exponent $\Phi$.  Including all data with 
$\beta \ge 0.378$ we get
\begin{equation}
\label{surf}
\bar{\beta}_{1,c}(\beta) = 0.54914 + 0.51736(12) [t -16.7(4) t^2 + 752(101) t^3
- 25015(6205) t^4]^{0.63002 \times 0.718}
\end{equation}
where $t=0.387721735-\beta$  with $\chi^2/$DOF $=1.23/2$. This result might 
be helpful in future studies of the surface transition in the BC1 model.

\section{Films with (sp,$+$) boundary conditions}
Finally we have simulated the BC1 model with  special boundary conditions at
one surface and $+$ boundary conditions at the other, where $+$ boundary 
conditions means that there are spins at $x_0 = L_0+1$ which are fixed to $+1$,
which is equivalent to $h_2=\beta$. We have simulated at $\beta=\beta_c$
and $\bar{\beta}_1=\bar{\beta}_{1,s} =0.549145$, which was our
preliminary estimate of $\bar{\beta}_{1,sp}$ at the time, when we started the 
simulations. We have measured the magnetisation profile. We have 
computed the Taylor expansion of the magnetisation profile 
in $\bar{\beta}_1$ around the simulation point $\bar{\beta}_{1,s}$ up
to the second order.
For each measurement  
we performed the following sequence of Monte Carlo updates:
One sweep with the local heat-bath update, a cluster-update, one  sweep
with the local heat-bath update,
and again a cluster-update. In these cluster-updates we 
flip the sign of the frozen boundary and all spins of the cluster    
attached to it. Performing this cluster update twice, we are back to $+$
boundary conditions at the surface.

Mainly, we have simulated lattices  with $L=4 \times L_0$.  
We have checked 
that at the level of our accuracy, deviations from the effectively 
two-dimensional thermodynamic limit $L/L_0 \rightarrow \infty$  are
negligible for this choice. To this end we have simulated 
$L=24$, $32$ and $48$ for 
the thickness $L_0=8$. The results obtained for $L=32$ and $48$
are consitent among each other and those for $L=24$ deviate only little.
For $L=4 \times L_0$, 
we have studied films of the thicknesses $L_0=6$, $8$, $12$, $16$,
$24$, $32$, $48$, $64$, $96$ and $128$. We have performed $10^9$,
$5 \times 10^8$, $10^8$, $10^8$, $10^8$, $6 \times 10^7$, $4 \times 10^7$,
$2.6 \times 10^7$, $1.5 \times 10^7$ and $7.7 \times 10^6$ measurements
for these lattice sizes, respectively.  In total these simulations took
a bit more than 2 years of CPU-time on a single core of a Quad-Core 
AMD Opteron(tm) Processor 2378 running at 2.4 GHz.

\subsection{The magnetization at the boundary}
\label{Mboundary}
First let us discuss the behavior of the magnetisation $m_1$ at the 
boundary  as a function of the thickness of the film.
In terms of the reduced free energy per area it 
is given by
\begin{eqnarray}
\frac{\partial f(L_0,t,h,t_1,h_1)}{\partial h_1}
&=& \frac{1}{L^2} \; \frac{1}{Z} 
\frac{\sum_{\{s\}} \exp(...+ h_1 \sum_{x_1,x_2} s_{(1,x_1,x_2)})}{\partial h_1}
\nonumber \\
&=&  \frac{1}{L^2} \left \langle \sum_{x_1,x_2}  s_{(1,x_1,x_2)} \right \rangle
=m_1\;\;.
\end{eqnarray}
Note that the non-singular contribution to the free energy from the first 
surface
does not feel the breaking of the symmetry by the second surface. Therefore
it is an even function of $h_1$ and does not contribute to the partial
derivative with respect to $h_1$.  Similar to eq.~(\ref{fsscaling}) the 
singular part of the reduced free energy per area behaves as
\begin{equation}
 f_{s}(L_0,t,h,t_1,h_1) = L_0^{-d+1} g(..., h_1 [L_0/l_{h_1}]^{y_{h_1}}) \;\;.
\end{equation}
Taking the partial derivative with respect to $h_1$ we get
\begin{eqnarray}
m_1 &=& \left . \frac{\partial f_{s}}{\partial h_1} \right |_{t=h=t_1=h_1=0} = 
 L_0^{-d+1} \left . 
\frac{\partial g(..., h_1 [L_0/l_{h_1}]^{y_{h_1}})}
 {\partial h_1}  \right |_{t=h=t_1=h_1=0}
\nonumber \\
&=&L_0^{-d+1} \left . 
g_{h_1}(0,0,0,0) 
\right |_{t=h=t_1=h_1=0}
 [L_0/l_{h_1}]^{y_{h_1}}
= c L_0^{-d+1+y_{h_1}} \;\;,
\end{eqnarray}
where $g_{h_1}$ denotes the partial derivative of $g$ with respect to
$h_1 [L_0/l_{h_1}]^{y_{h_1}}$. 

We have fitted the magnetisation at the boundary with the ansaetze
\begin{equation}
\label{am1}
 m_1(L_0) = c (L_0+L_s)^x 
\end{equation}
with the free parameters $c$, $L_s$ and $x$ and 
\begin{equation}
\label{am2}
 m_1(L_0) = c (L_0+L_s)^x \times (1 + d (L_0+L_s)^{-2}) \;\;,
\end{equation}
where we have included corrections $\propto L_0^{-1}$ and $\propto L_0^{-2}$, 
respectively.
Using the Taylor-expansion, we have evaluated $m_1$  at
$\bar{\beta}_1=0.54914$ and in order to estimate the effect of the 
uncertainty of our estimate of $\bar{\beta}_{1,sp}$ also at 
$\bar{\beta}_1=0.54916$.
Results of fits with the ansaetze~(\ref{am1},\ref{am2}) using the 
data for $\bar{\beta}_1=0.54914$ are given in table \ref{tabmag}.

\begin{table}
\caption{\sl  \label{tabmag} Fitting the surface magnetisation $m_1$
at $\bar{\beta}_1=0.54914$. 
}
\begin{center}
\begin{tabular}{cccccc}
\hline
Ansatz & $L_{0,min}$  & $x$      &  $L_s$   & $d$ &  $\chi^2/$DOF \\
\hline
\ref{am1} &  8  &  --0.35303(7)  & 0.908(3) &         & 31.87/6  \\
\ref{am1} & 12  &  --0.35341(12) & 0.938(8) &         & 14.37/5  \\
\ref{am1} & 16  &  --0.35377(16) & 0.975(14)&         & 3.13/4   \\
\ref{am2} &  6  &  --0.35367(14) & 0.988(14)& 0.12(2) & 10.46/6 \\
\ref{am2} &  8  &  --0.35396(21) & 1.029(27)& 0.19(5) & 6.97/5 \\
\hline
\end{tabular}
\end{center}
\end{table}
Repeating the fit with the ansatz~(\ref{am1}) and $L_{0,min}=16$ for the
data taken at $\bar{\beta}_1=0.54916$ we get $x=-0.35280(16)$, $L_s=0.935(13)$ and 
$\chi^2/$DOF $=2.98/4$.  Fitting the data for $\bar{\beta}_1=0.54916$
with the ansatz~(\ref{am2}) and $L_{0,min}=8$ we get $x=-0.35288(18)$, 
$L_s=0.965(22)$, $d=0.12(3)$ and $\chi^2/$DOF $=6.08/5$.  We see that the error
of the exponent $x$ as well as that of $L_s$ is dominated by the uncertainty of 
$\bar{\beta}_{1,sp}$. As our final estimate we quote $x=-0.3539(13)$ and hence
\begin{equation}
y_{h_1} = 1.6461(13)  
\end{equation}
which is fully consistent with but less precise than  the estimate obtained above,
eq.~(\ref{yshfinal}). For the effective shift in the thickness of the lattice 
we quote $L_s=1.0(1)$.   Note that $L_s$ should be equal to the sum of the extrapolation
lengths at the two surfaces. For $+$ boundary conditions we find in ref. \cite{H-11} 
the result
$l_{ex} =0.96(2)$ for the $+$ boundary conditions. Hence $l_{ex,sp} \approx 0$ for
the BC1 model. This is consistent with our observation above that for the BC1 model
corrections $\propto L_0^{-1}$ have a small amplitude.

We also have performed fits with the ansaetze~(\ref{am1},\ref{am2}), where 
$\bar{\beta}_{1,sp}$ is a free parameter. To this end we have used the Taylor series
of the surface magnetisation $m_1$ in $\bar{\beta}_1$ around the simulation point.
Such fits suggest $\bar{\beta}_{1,sp}=0.54916(4)$, where the error
is larger than that of our final estimate~(\ref{finalsp}). Correspondingly, 
also the estimate of $y_{h_1}$ cannot be improved this way.

\subsection{The universal scaling function of the magnetisation profile}
The  magnetisation profile at the critical point behaves as
\cite{BinderS,Diehl86,Diehl97}
\begin{equation}
\label{magfinite}
 m(z) = c \; L_0^{-\beta/\nu} \; \psi(z/L_0)
\end{equation}
where $z$ gives the distance from the boundary. The universal function 
$\psi$ depends on the surface universality classes of the boundary 
conditions at the two surfaces of the film. In the neighborhood of the 
surface, the magnetisation follows a power law
\begin{equation}
 m(z) = a z^{-\beta/\nu+d-1-y_{h_1}} \;\;.
\end{equation}
Note that the exponent $-\beta/\nu+d-1-y_{h_1} =-0.1646(7)$ is negative
and therefore the magnetisation is actually decreasing with increasing 
distance from the surface. This is in contrast to ordinary boundary conditions, 
where $y_{h_1}= 0.7249(6)$ \cite{H-11}, and therefore the magnetisation is
increasing with the distance.
Note that from scaling relations it follows that $\beta/\nu=(1+\eta)/2$, where
$\eta=0.03627(10)$ for the three-dimensional Ising universality class
\cite{mycritical}. 

\begin{figure}[tp]
\includegraphics[width=13.5cm]{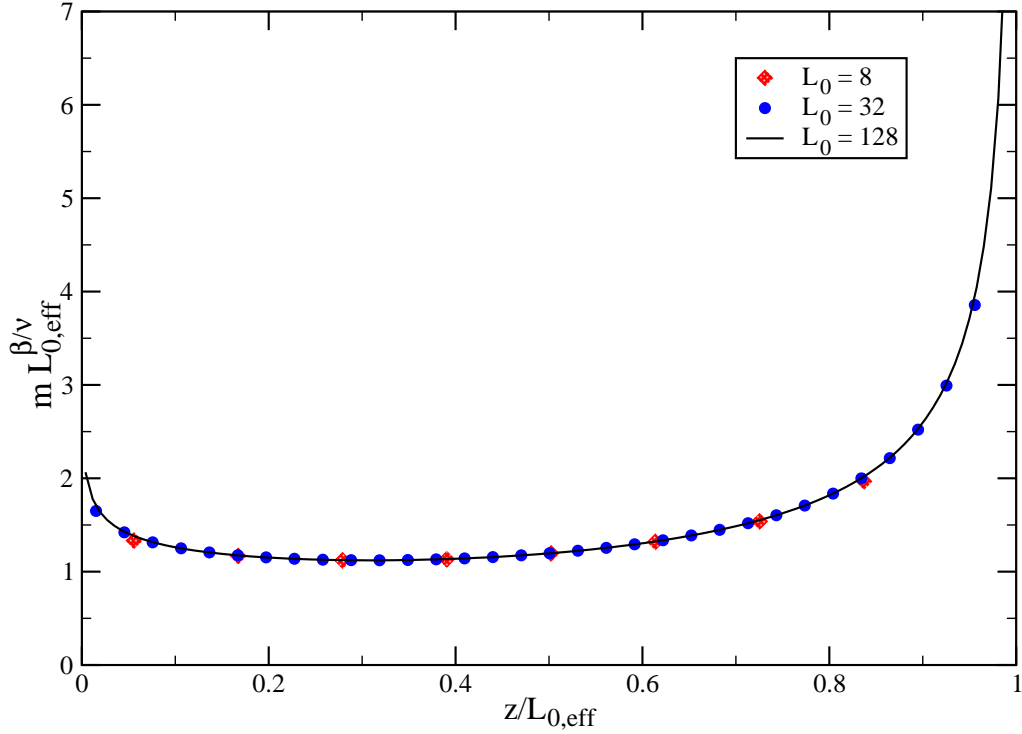}
\caption{\label{phiplot}
We plot $m(z) L_{0,eff}^{\beta/\nu}$ as a function of $z/L_{0,eff}$, 
where $z$ is the distance from the boundary with special boundary
conditions. The data are taken at the bulk critical point. At the first boundary 
special boundary conditions and at the second $+$ boundary conditions are 
applied. 
We give data for the thicknesses $L_0=8$, $32$ and $128$. The line given for $L_0=128$
is obtained from a linear interpolation of the data points.  We abstained from giving 
data for more thicknesses to keep the figure readable. Notice that the statistical
errors are smaller than the symbols and the line shown.
For a detailed discussion see the text. Colour online.
}
\end{figure}

Using the Taylor-expansion up to
second order we have computed the magnetisation profile for $\bar{\beta}_1=0.54914$,
which is our estimate of $\bar{\beta}_{1,sp}$. 
In figure \ref{phiplot} we have plotted  $m(z) L_{0,eff}^{\beta/\nu}$ as a function
of $z/L_{0,eff}$. In order to take corrections $\propto L_0^{-1}$ into account,
we have replaced in eq.~(\ref{magfinite}) the thickness $L_0$ by the effective one 
$L_{0,eff}=l_{ex,sp}+l_{ex,+}$. The distance from the boundary is given by 
$z=x_0-1/2+l_{ex,sp}$. As discussed above, the extrapolation lengths take the 
values $l_{ex,+}=0.96$ and $l_{ex,sp}=0$. 
We see that the data for the three different thicknesses fall nicely on top 
of each other. The same holds for thicknesses not plotted here. Therefore the curve 
obtained from $L_0=128$ should be a good approximation of the universal scaling 
function $\psi(z/L_0)$. We observe that, as discussed above, for small $z$ the 
magnetisation is indeed decreasing with increasing $z$.  At $z/L_0 \approx 0.31$ a 
very shallow minimum is reached.

\section{Summary and Conclusions}
In this work we have studied the special point in the phase diagram of 
a semi-infinite system. We have accurately estimated the surface RG-exponents 
$y_{t_1}$ and $y_{h_1}$ that govern, along with the bulk RG-exponents $y_t$ and $y_h$,
the behavior of surface quantities at the special point. To this end, we have simulated 
the improved Blume-Capel model with two different surface interactions and the 
spin-1/2 Ising model on the simple cubic lattice. 
We have determined the surface RG-exponents
from a finite size scaling analysis at the special point. In the presence of surfaces
typically corrections $\propto L_0^{-1}$ appear, where $L_0$ is the linear extension 
of the finite system. Fitting data it is difficult to disentangle these from
the leading bulk corrections $\propto L_0^{-\omega}$, where $\omega=0.832(6)$ 
\cite{mycritical}. Therefore it is helpful that in the improved Blume-Capel model
the amplitude of the leading correction to scaling vanishes. 

In table \ref{expospecial} we compare our results with those obtained in previous 
Monte Carlo (MC) studies and computed by field theoretic methods. In previous
MC studies only the spin-1/2 Ising model on the simple cubic lattice had been 
simulated. Only in ref. \cite{DeBlNi05}  results for the RG-exponents are quoted. 
However all authors give results for the exponents $\beta_1$ and most for 
$\Phi$. 
In table \ref{expospecial} we have converted these estimates by using the scaling
relations $\beta_1 = \nu (2 - y_{h_1})$ and $\Phi = \nu y_{t_1}$, where we 
have used $\nu=0.63002$.
In the case of $y_{h_1}$ we observe that the estimates of refs. \cite{RuWa95,DeBlNi05}
deviate by $9$ and $6.6$ times the combined error from our result. 
Notice that such deviations can be caused by corrections that are not taken into account
in the ansatz. For a general discussion of this problem see section \ref{corrections}.
The fits performed in ref. \cite{DeBlNi05} are in particular prone to this problem,
since in the ansatz corrections $\propto L_0^{-3}$ are included while e.g. corrections
$\propto L_0^{-2 \omega}$ are missing. 
In the present work, we tried to estimate such systematical
errors by comparing the results obtained by fitting with ansaetze of a varying number 
of correction terms.
In the case of $y_{t_1}$ we see a quite good agreement 
with the results of refs. \cite{RuDuWaWu93,DeBlNi05}. In contrast, the estimates
of refs. \cite{BiLa84,LaBi90,VeRoFa92} are clearly larger than ours.

The surface critical exponents for the special point have been computed up to 
second order in the $\epsilon$-expansion. Direct results have been obtained 
for the exponents $\eta_{||}$ \cite{Reeve} and $\Phi$ \cite{DiDi81b}. Starting 
from these two exponents and the bulk exponents, the remaining ones can be 
computed by using scaling relations. Naively inserting $\epsilon=1$ one gets
$\eta_{||}=-0.30$ and $\Phi=0.68$ for three dimensions; see table VI of ref. 
\cite{Diehl86}. In table
\ref{expospecial} we give $y_{h_1}= (d-\eta_{||})/2$ and $y_{t_1} = \Phi/\nu$, 
where we have used $\nu=0.63002$. The result for $y_{h_1}$ is in quite good 
agreement with those from Monte Carlo simulations. In contrast, the estimate 
for $y_{t_1}$ is by far too large.  The authors of ref. \cite{DiSh98} have 
pointed out that by resummation of the series, e.g. by Pad{\'e} approximants
this discrepancy can be resolved. The authors of ref. \cite{DiSh98} have computed 
the surface exponents in a two-loop calculation in three-dimensions fixed.
The numerical estimates quoted by the authors are obtained by using Pad{\'e} 
approximants.
The numbers given in table \ref{expospecial} are again obtained by using 
$y_{h_1}= (d-\eta_{||})/2$ and $y_{t_1} = \Phi/\nu$, where $\nu=0.63002$. 
The results for $y_{h_1}$ and in particular for $y_{t_1}$ deviate clearly
from those obtained from Monte Carlo simulations.

\begin{table}
\caption{\sl \label{expospecial}
Results for the surface RG-exponents of the
special phase transition in the three-dimensional Ising bulk universality class.
}
\begin{center}
\begin{tabular}{cclll}
\hline
 Ref.             &  year   & Method  &  $y_{h_1}$    &  $y_{t_1}$  \\
\hline
\cite{BiLa84}   & 1984  &  MC   & 1.72(4)  & 0.89(6) \\
\cite{LaBi90}    & 1990  &  MC   & 1.71(3)  & 0.94(6) \\
\cite{VeRoFa92}  & 1992  &  MC   & 1.65    & 1.17    \\
\cite{RuDuWaWu93}& 1992  &  MC  & 1.624(8) & 0.732(24) \\
\cite{RuWa95}    & 1995  & MC   & 1.623(2) &         \\
\cite{SePl98}     & 1998 & MC   & 1.635(16) &          \\
\cite{DeBlNi05}   & 2005 & MC  & 1.636(1)  & 0.715(1) \\
this work         & 2011 & MC  & 1.6465(6) & 0.718(2) \\
\cite{Reeve,DiDi81b,Diehl86} & 1981 &$\epsilon$-exp, naive &1.65 &  1.08  \\
 \cite{DiDi81b,DiSh98}         & 1998  &$\epsilon$-exp, resummed &     & 0.752 \\
 \cite{DiSh98}    & 1994 & 3D-exp, resummed & 1.583   & 0.856 \\
\hline
\end{tabular}
\end{center}
\end{table}

In addition to the estimates of the RG-exponents, our finite size analysis 
provides us with accurate estimates of the surface coupling $\bar{\beta}_{1,sp}$
at the special point for the three models that we have simulated. These 
estimates can be used in future studies of thin films. In particular, we 
intend to study the thermodynamic Casimir force in thin films in the neighborhood 
of the special point.

Furthermore we have studied the behavior of the surface transition in the 
neighborhood of the special point. Our numerical results follow the theoretical
expectations. However the special point cannot be located accurately with such 
an approach.

Finally we have simulated a film with symmetry breaking boundary conditions at
one surface and special boundary conditions at the other. The behavior of the 
magnetisation at the surface with  special boundary conditions follows a power law.
Its exponent can be expressed in terms of the RG-exponent $y_{h_1}$.  The analysis
of the data fully confirms our estimate of $y_{h_1}$ given in table \ref{expospecial}.
The magnetisation profile follows a universal function. This theoretical expectation
is fully confirmed by the nice collapse of data that we observe for a large range
of thicknesses $L_0$ of the film. An interesting feature of the magnetisation 
profile is that at the surface with special boundary conditions, which do not 
break the symmetry, the magnetisation is decreasing with increasing distance from the 
surface.

\section{Acknowledgements}
This work was supported by the DFG under the grant No HA 3150/2-2.

\end{document}